\pdfoutput=1
\documentclass[twoside,twocolumn]{article}

\immediate\write18{/usr/local/bin:/Users/nadavamir/Dropbox/Meditation_EEG/AB/crop_all.sh}

\usepackage{enumitem}
\usepackage{amsmath}
\usepackage{mathabx}
\usepackage{subcaption}

\usepackage{graphicx}
\graphicspath{{./figures/}}

\usepackage{epstopdf}
\usepackage{caption}

\usepackage{blindtext} 

\usepackage[sc]{mathpazo} 
\usepackage[T1]{fontenc} 
\usepackage{microtype} 

\usepackage[english]{babel} 

\usepackage[hmarginratio=1:1,top=32mm,columnsep=20pt]{geometry} 

\usepackage{abstract} 

\usepackage{titlesec} 
\renewcommand\thesection{\Roman{section}} 
\renewcommand\thesubsection{\roman{subsection}} 
\titleformat{\section}[block]{\large\scshape\centering}{\thesection.}{1em}{} 
\titleformat{\subsection}[block]{\large}{\thesubsection.}{1em}{} 

\usepackage{fancyhdr} 
\usepackage{titling} 

\usepackage{hyperref} 

\pretitle{\begin{center}\Huge\bfseries} 
\posttitle{\end{center}} 
\title{A Simple Model of Attentional Blink} 
\author{
\textsc{Nadav Amir} \\[1ex] 
\normalsize  The Edmond and Lily Safra Center for Brain Sciences\\ \normalsize The Hebrew University of Jerusalem\\
\normalsize \href{mailto:nadav.amir2@mail.huji.ac.il}{nadav.amir2@mail.huji.ac.il} 
\and 
\textsc{Israel Nelken} \\[1ex] 
\normalsize Department of Neurobiology, Institute for Life Sciences \& \\\normalsize The Edmond and Lily Safra Center for Brain Sciences  \\ %
\normalsize The Hebrew University of Jerusalem\\
\normalsize \href{mailto:israel@cc.huji.ac.il}{israel@cc.huji.ac.il} 
 \and 
 \textsc{Naftali Tishby} \\[1ex] 
 \normalsize The Rachel and Selim Benin School of Computer Science and Engineering \& 
 \\\normalsize The Edmond and Lily Safra Center for Brain Sciences  \\ 
 \normalsize The Hebrew University of Jerusalem\\
 \normalsize \href{mailto:tishby@cs.huji.ac.il}{tishby@cs.huji.ac.il} 
}
\date{\today} 
\renewcommand{\maketitlehookd}{%
\begin{abstract}
\noindent The attentional blink (AB) effect is the reduced ability of subjects to report a second target stimuli (T2) among a rapidly presented series of non-target stimuli, when it appears within a time window of about 200-500 ms after a first target (T1). We present a simple dynamical systems model explaining the AB as resulting from the temporal response dynamics of a stochastic, linear system with threshold, whose output represents the amount of attentional resources allocated to the incoming sensory stimuli. The model postulates that the available attention capacity is limited by activity of the default mode network (DMN), a correlated set of brain regions related to task irrelevant processing which is known to exhibit reduced activation following mental training such as mindfulness meditation. The model provides a parsimonious account relating key findings from the AB, DMN and meditation research literature, and suggests some new testable predictions. 
\end{abstract}
}

\begin{document}

\maketitle


\section{Introduction}
Questions regarding the characteristics, and limitations, of attention allocation over time have attracted considerable interest over the last decades (see \cite{Shapiro2001} for an overview). More recently, it has been shown that intensive mental training, in the form of mindfulness meditation, can modulate the control of attention allocation over time \cite{Slagter2007,Lutz2009a,VanVugt2014}, yet the computational and neural mechanisms underlying this process remain unclear. A possible clue lies in the recent deluge of studies concerning the so called default mode network (DMN) (see \cite{Raichle2001,Raichle2015} for reviews): an anatomically defined, interconnected system of cortical regions which are preferentially activated when individuals are involved in self referential processing and mind wandering, rather than paying attention to the external environment \cite{Mason2007,Christoff2009,Buckner2008}. Various meditation practices have been shown to be associated with reduced activity levels of the DMN \cite{Garrison2015,Brewer2011,Farb2007,Berkovich-Ohana2016}, as well as improved sustained \cite{Pagnoni2012} and selective \cite{Lutz2009a} attention capabilities.
\par
Here we describe a simple dynamical systems model suggesting that changes in DMN activity levels can modulate attentional capacity. The proposed mechanism is quite general in nature but we demonstrate it using the well-known attentional blink (AB) paradigm \cite{Raymond1992,Broadbent1987}, which is the reduced ability (``blinking") of subjects asked to report a second target stimuli in a rapid serial visual presentation (RSVP) task, when it appears within a time window of about 200-500 ms after the first one. This AB effect has played a central role in studying the limits of humans' ability to allocate attention over time (for reviews see \cite{Dux2009rev,Martens2010}) and more recently has also been used to study the effects of mental training (in the form of mindfulness meditation) on the temporal distribution of attentional resources \cite{Slagter2007,Lutz2009a,VanVugt2014}. 
While several theoretical accounts have been suggested over the last decade for explaining the AB (see discussion below), the model proposed here has the advantage of being simple and parsimonious in parameters, while accounting for many empirical findings relating the AB, DMN and mental training research literature. Additionally, it provides testable hypotheses relating changes in DMN activity levels with performance in attention demanding tasks and associated ERP amplitudes.
\section{Methods}
\subsection{Basic description of the model}
Formally, we model the ``attentional channel'' as a linear time-invariant (LTI) system \cite{Chen1984} with additive Gaussian noise and an upper response threshold (Eq.~\ref{eq:model}). The system's output, or response, $y(t)$, corresponds to the amount of attentional resources currently used by the brain. When this response reaches a predetermined \textit{blinking threshold}, $y(t)=y_B$, the attentional resources are exhausted and the system cannot respond to incoming stimuli for a short ``refractory'' period, resulting in decreased identification of the subsequent target stimuli - the AB effect. 
\par
The system's response consists of deterministic and stochastic components. The former is given by a convolution of the input stimuli's  \textit{cognitive representation} $u_{c}(t)$ (see next subsection), with the system's impulse response function $h(t)$, which describes the time course of the system's response to a pulse shaped input. The stochastic part of the response is given by the additive Gaussian random variable $n_{DMN}(t)\sim\mathcal{N}(\mu_{DMN},\sigma^2_{DMN})$, with $\mu_{DMN}$ representing the DMN activity baseline (mean) and $\sigma_{DMN}$ its fluctuations (standard deviation). Conceptually, DMN activity is represented by the model as task irrelevant noise loading the attentional system and thus modulating blinking propensity. 
\par
To summarize, we present a model describing the temporal capture of attention using a dynamical system  described by Eq. ~\ref{eq:model}, which is specified by the cognitive representation of the input stimuli $u_{c}(t)$, the blinking threshold $y_B$, the impulse response function $h(t)$ and the DMN noise $n_{DMN}$. 
\begin{equation}
y(t)=\min 
\begin{cases} 
u_{c}(t)*h(t)+n_{DMN}(t)\\
y_B
\end{cases}
\label{eq:model}
\end{equation} 
\subsection{Cognitive representation of input stimuli}
In line with other two-stage accounts of the AB (e.g. \cite{Chun1995,Broadbent1987} and see discussion below), the model assumes that sensory stimuli undergo an initial \textit{cognitive representation} stage before they can be further processed, or consolidated into working memory. The input sensory stimuli stream is represented by a discrete (10Hz) signal of impulses $u_{s}(t)$, with amplitudes corresponding to stimulus saliency. Target stimuli are represented as unit impulses and non-targets as zeros, i.e. $u_{s}(t)=1$ if one of the target stimuli (T1 or T2) appears at time $t$ and $u_{s}(t)=0$ otherwise. This formulation can be easily adjusted to account for various target/distractor saliency relationships by modulating the impulse amplitudes of the different stimuli (see discussion section below). Concretely, the representation stage is implemented by resampling the sensory stimulus signal $u_{s}(t)$ at a higher sampling rate (1KHz), and clipping it at a threshold value $u_{c}(t)=1$ (not to be confused with the blinking threshold $y_B$ in Eq.~\ref{eq:model}). The resulting cognitive representation signal $u_{c}(t)$, is then used as input to the attentional system described by Eq.~\ref{eq:model}. Conceptually, resampling represents the temporal resolution of the attentional system and clipping corresponds to sub-linear combination of temporally close sensory stimuli. Importantly, the sub-linearity is most pronounced when the two target stimuli appear successively, in which case their combined cognitive representation signal exhibits maximal overlap and clipping. As discussed below, this sub-linear summation of temporally proximal target stimuli can explain various effects related to the somewhat counterintuitive finding that blinking is significantly attenuated when T2 appears directly after T1 - the so called ``lag-1 sparing'' effect \cite{Potter1998}. 
\subsection{DMN activity as attentional noise}   
A central feature of the model is its interpretation of DMN activity as stochastic noise in the attentional system. This is formally described by the additive noise term $n_{DMN}(t)$ in Eq.~\ref{eq:model} which is a Gaussian random variable, $n_{DMN}(t)\sim\mathcal{N}(\mu_{DMN},\sigma^2_{DMN})$, with mean and variance corresponding to the DMN's baseline activity and moment-by-moment fluctuation levels respectively. Conceptually, the mean DMN activity level represents the underlying mental state (e.g. mind wandering vs task engagement), whereas the variance represents momentary attentional fluctuations around the baseline level. A lower DMN activity level results in more attentional resources available for task performance and less probability of crossing the blinking threshold. This mechanism enables the model to relate DMN activity and AB task performance in a quantitative way. The recognition probability of T2 can be directly computed from the \textit{blinking gap}, defined as the difference between $y_B$ (the blinking threshold), and the maximal value of $y(t)$ during the system's excitation by the stimulus inputs.
\subsection{The impulse response function of the attentional system}  
The impulse response function $h(t)$, determines the temporal profile of attentional resource allocation. It is modeled here using the Gamma (or Erlang) distribution function (Eq.~\ref{eq:irf}): 
\begin{equation}
h(t)= \begin{cases} st^{(n-1)}e^{-t/\tau} & t>0\\
0 & t\leq0
\end{cases}
\label{eq:irf}
\end{equation} 
Here $n$ and $\tau$ are the shape and scale parameters which fix the half duration and maximum response time of $h(t)$ (Fig.~\ref{fig:impulse_response}) and the scaling factor $s$ controls the system's gain. While the precise form of $h(t)$ is not of major significance, the Gamma distribution function is known to provide a reasonably good fit for attention related responses such as pupillary dilation (a correlate of attentional effort) \cite{Hoeks1993,Wierda2012} , temporal sensitivity in the visual system \cite{Watson1986} and even Blood-oxygen-level dependent (BOLD) signal hemodynamic responses \cite{Boynton2012}.
\subsection{The P3b brain potential}
The P3b is a positive ERP, peaking at around 300ms, associated with updating of working memory \cite{Polich2007,Polich2006,Donchin1988} and attentional resources allocation \cite{Wickens1983,JohnsonR.1988}. To interpret the P3b amplitude in terms of the model's parameters, we first define the \textit{Resource Allocation Index} (RAI), as the ratio between attentional resources allocated to the stimuli under consideration and the total resources available (Eq.~\ref{eq:rai}).
\begin{equation}
RAI=\frac{\max_{t}{y(t)}-\mu_{DMN}}{y_{B}-\mu_{DMN}}
\label{eq:rai}
\end{equation} 
The maximum is taken over the time period during which the system is responding to the stimuli under consideration. Note that the RAI is a dimensionless number taking values between $0$, corresponding to no excitation of the attentional system and $1$, representing maximal excitation or exhaustion of attentional resources. Since P3b amplitude ranges typically differ between subjects, even when performing the same task, we define the model correlate of the P3b amplitude as the RAI divided by the subject specific DMN activity standard deviation $\sigma_{DMN}$, which serves as a natural signal to noise scaling factor in this context (Eq.~\ref{eq:P3b_model}).
\begin{equation}
P3b=\frac{RAI}{\sigma_{DMN}}
\label{eq:P3b_model}
\end{equation} 
 How are changes in DMN activity levels reflected in the P3b amplitude defined by equation \ref{eq:P3b_model}? For non-blinking trials, i.e. when $RAI<1$, the $RAI$ and thus the P3b amplitude increases as $\mu_{DMN}$ is decreased and $\sigma_{DMN}$ remains fixed. However, when both $\mu_{DMN}$ and $\sigma_{DMN}$ are decreased, the expected P3b amplitude may increase or decrease depending on their specific values, since reducing $\sigma_{DMN}$ reduces both the numerator in equation \ref{eq:rai} and the denominator in \ref{eq:P3b_model}. For blinking trials, i.e. when $RAI=1$, the situation is simpler since in this case $P3b=\sigma_{DMN}^{-1}$ regardless of $\mu_{DMN}$'s value. 
\begin{figure}[!htbp]
\includegraphics[width=0.5\textwidth]{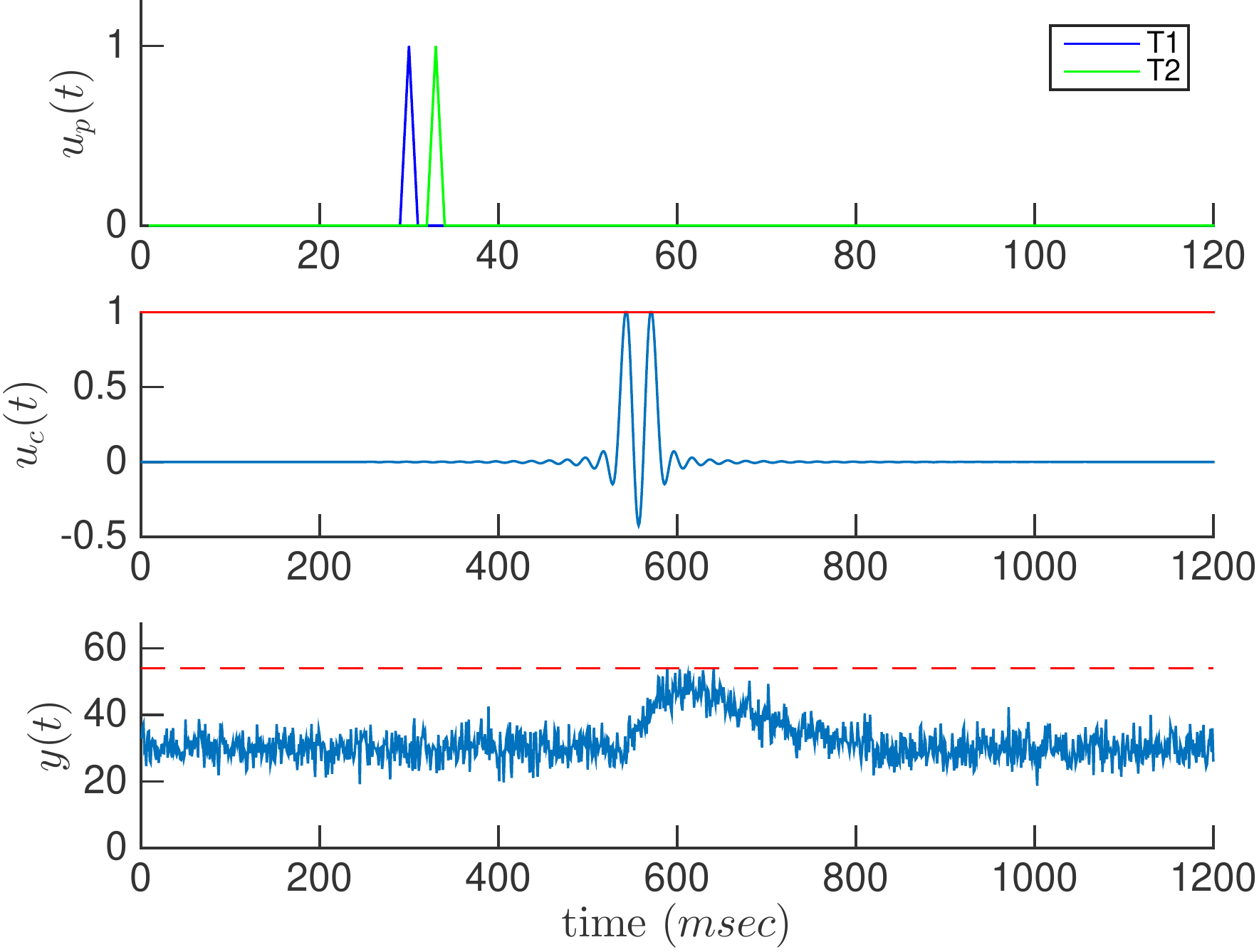}
\includegraphics[width=0.5\textwidth]{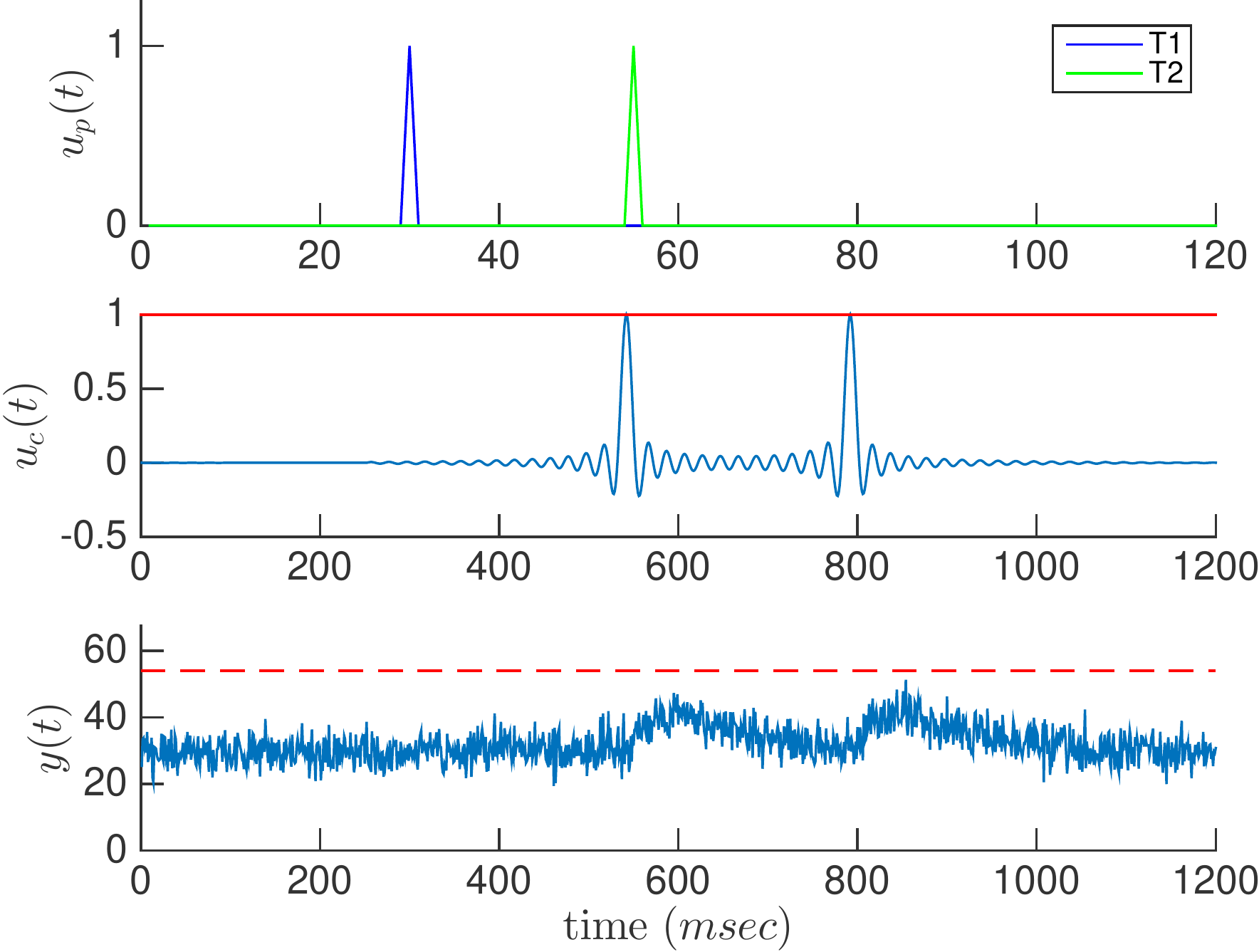}
\caption{{\bf Attentional Blink model}
Top: When T2 (in green) appears close (but not directly, see Fig.~\ref{fig:lag1sparing}) after T1 (blue), the  blinking threshold (red line) is reached and blinking occurs. Bottom: When the time interval between T1 and T2 is longer, both targets can be processed without causing the system to reach the blinking threshold.}
\label{fig:AB_model}
\end{figure}  
\begin{figure}[!htbp]
\includegraphics[width=0.5\textwidth]{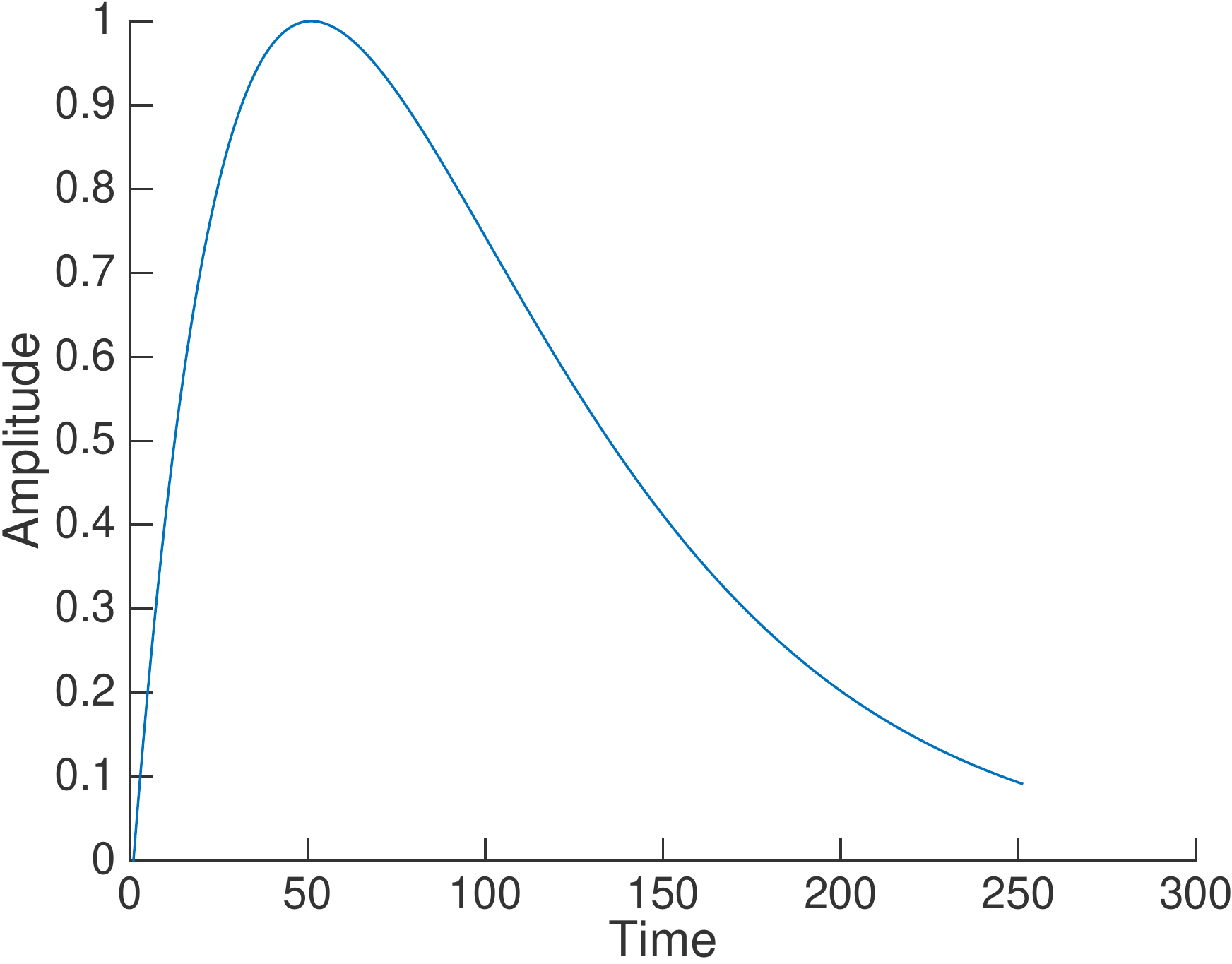}
\caption{{\bf Impulse response function}} 
The temporal response of the attentional system to an impulse stimuli follows the Gamma distribution function (Eq.~\ref{eq:irf}). The values of the shape and scale parameters used here are $n=1$ and $\tau=0.05$ respectively.
\label{fig:impulse_response}
\end{figure} 
\section{Results}
\begin{figure}[!htbp]
\includegraphics[width=0.5\textwidth]{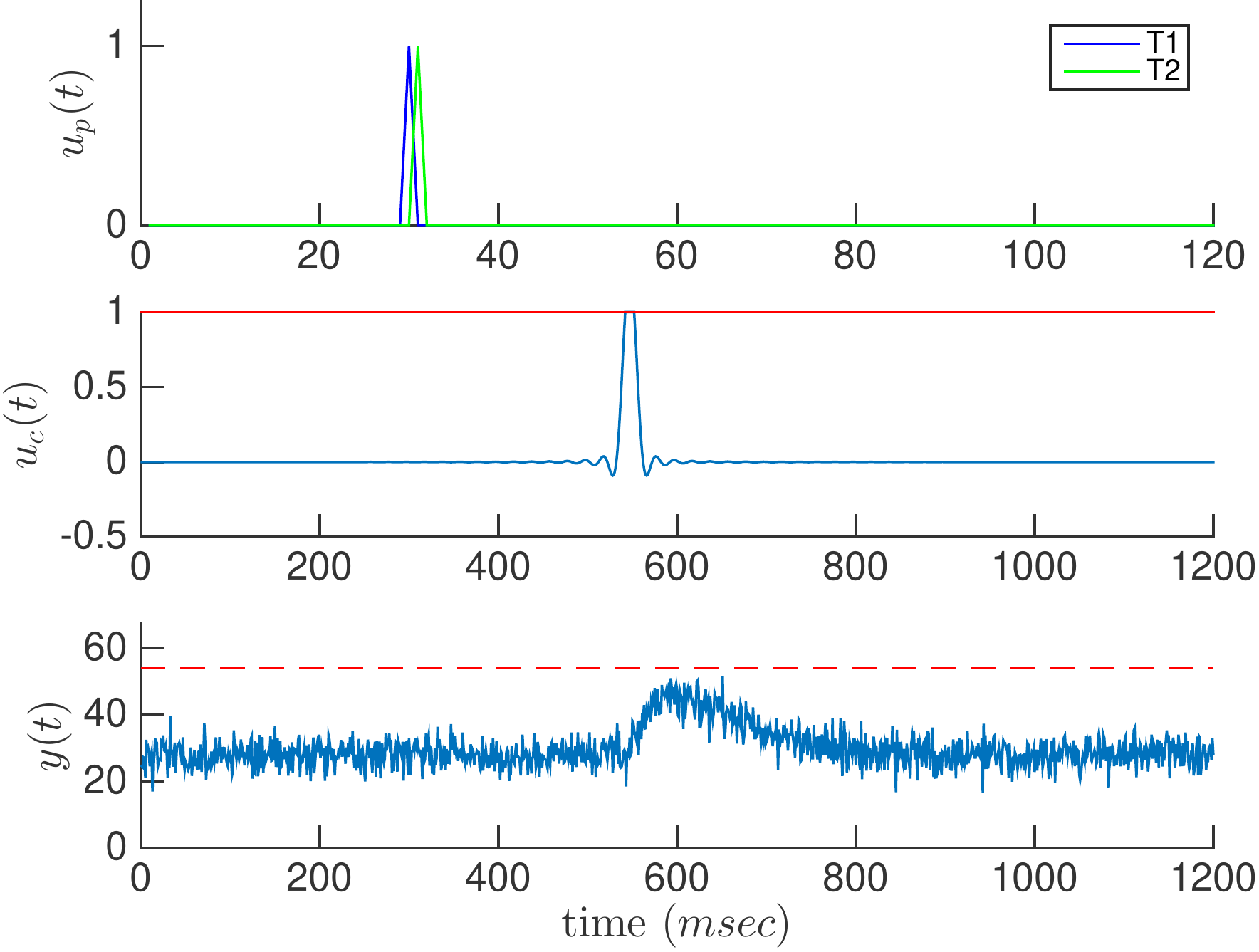} 
\includegraphics[width=0.5\textwidth]{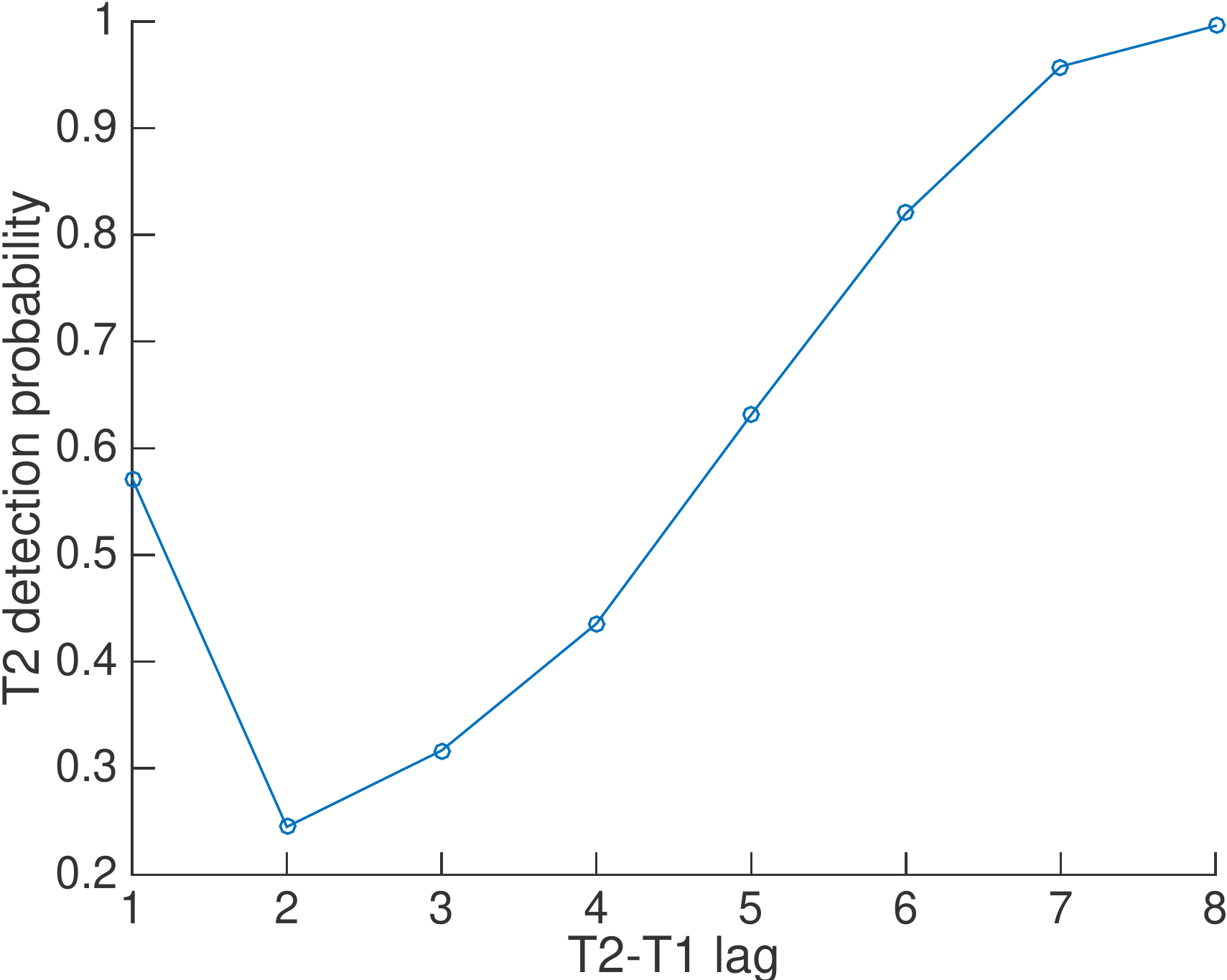}
\caption{{\bf Lag-1 sparing}
Top: When T2 (green) appears immediately after T1 (blue), their larger overlap at the cognitive representation stage results in more clipping which in turn reduces the maximal response of the system and decreases blinking probability. Bottom: Probability of T2 detection as a function of the time lag between T1 and T2. This u-shaped profile is typical in AB experiments (\cite{Raymond1992}).}
\label{fig:lag1sparing}
\end{figure}

We first tested whether the model can reproduce the basic AB effect, namely a reduction in detection accuracy of T2 within a time window of approximately 200-500 ms after T1 presentation. Due to the resampling of stimuli at the cognitive representation stage, and the finite rise and fall time of the system's impulse response function (Fig.~\ref{fig:impulse_response}), two targets which appear close to each other will have overlapping responses. This overlap can cause the system's response to the combined signal to reach the blinking threshold, even though the individual response to each one of the targets does not do so (Fig.~\ref{fig:AB_model}, top). As the time interval between T1 and T2 increases, their overlap decreases and the system's response to the combined signal no longer reaches the blinking threshold (Fig.~\ref{fig:AB_model}, bottom). Importantly, when T2 appears immediately after T1, the combined response may not cross the blinking threshold since in this case a significant portion of the overlap between targets is clipped at the cognitive representation stage (Fig.~\ref{fig:lag1sparing}), resulting in a sub-linearly combined signal.
\par
\begin{figure}[htbp]
	\includegraphics[width=0.5\textwidth]{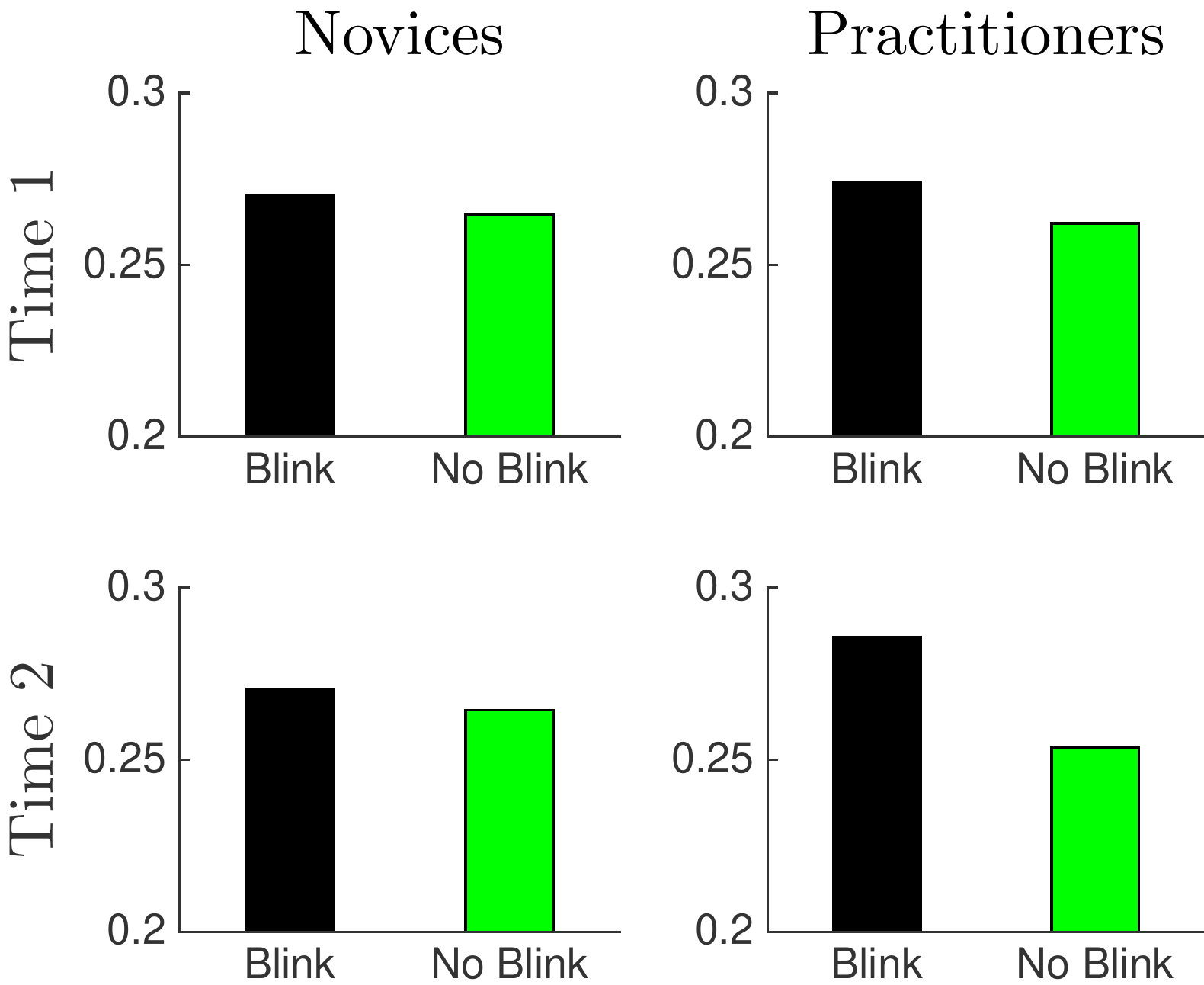} 
	\includegraphics[width=0.5\textwidth]{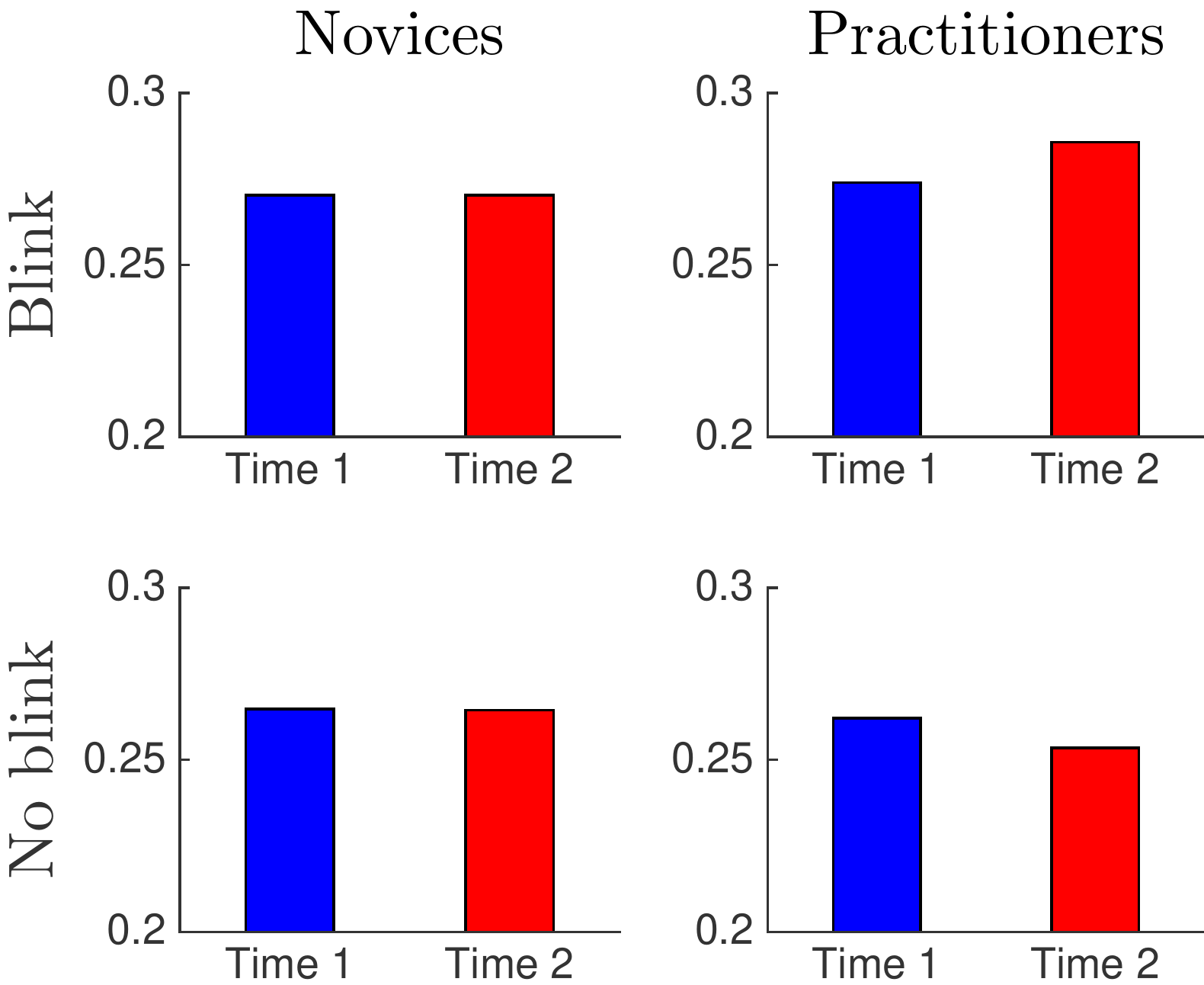}	
	\caption{{\bf Effects of mental training on T1 elicited P3b}	
		Top: Modeled T1 evoked P3b amplitude as a function of T2 accuracy: no-blink vs. blink, session: Time 1 (before the meditation retreat) vs. Time 2 (after the retreat), and group (practitioners vs. novices). Meditation practitioners show a greater reduction in T1 evoked P3b amplitude compared to novices in no-blink vs blink trials at time 2 vs time 1. Bottom: Selective reduction in T1 evoked P3b amplitude in no-blink trials in the practitioner group. DMN parameter levels $(\mu_{DMN},\sigma_{DMN})$, for  practitioners: $(3.65,28)$ at time 1 and $(3.5,25)$ at time 2. For novices: $(3.7,30)$ at both times. These results, and choice of colors, follow figure 3 in \cite{Slagter2007}.The larger T1 evoked P3b amplitude for practitioners at time 1, for blinking vs. non-blinking trials and at time 2 vs. time 1, for blinking trials are novel predictions of the model.}	
	\label{fig:meditationEffect}
\end{figure}

\begin{figure}[!htbp]
\begin{subfigure}{.25\textwidth}
	\centering \includegraphics[width=1\linewidth]{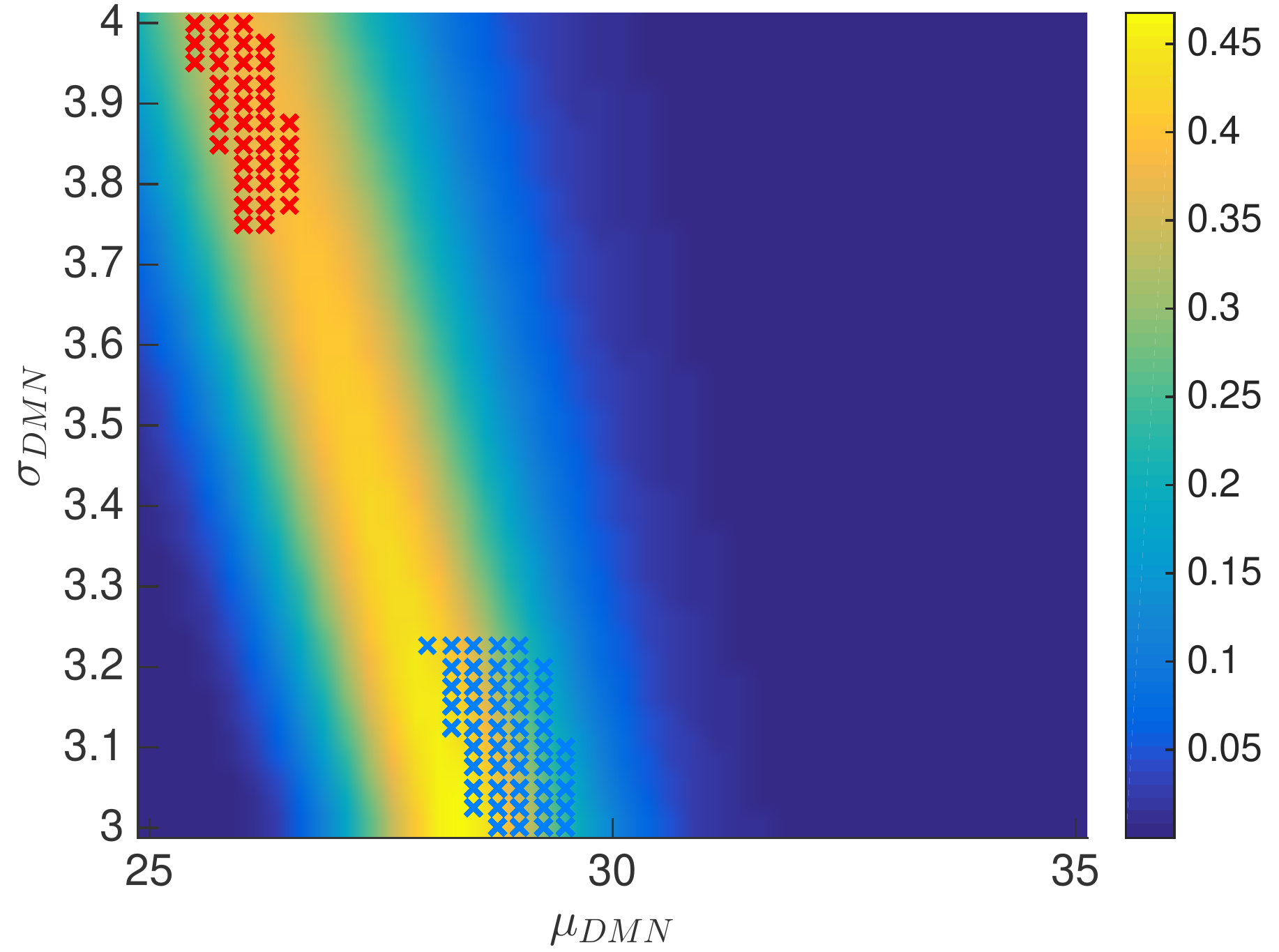}
\end{subfigure}%
\begin{subfigure}{.25\textwidth}
	\centering \includegraphics[width=1\linewidth]{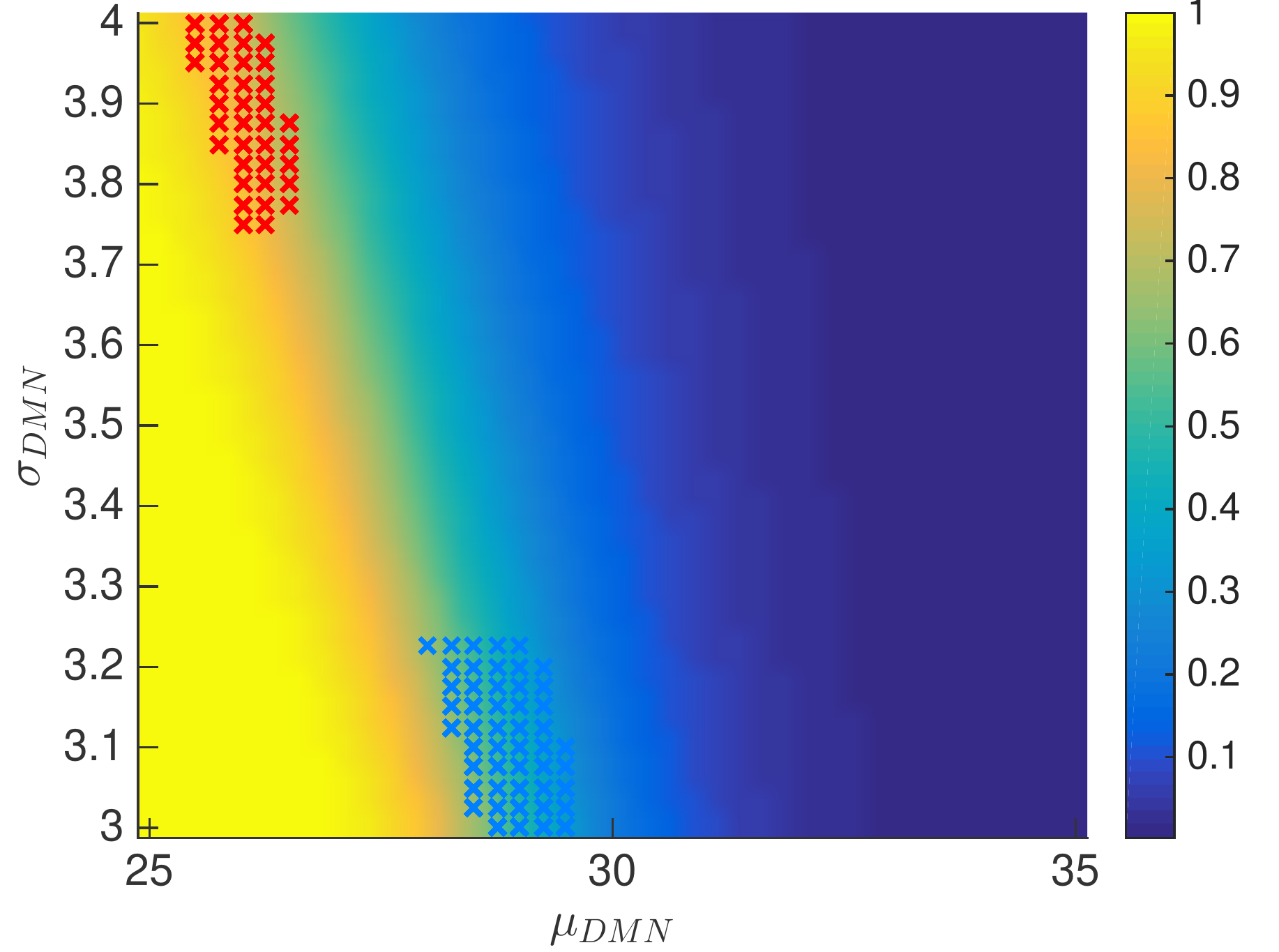}
\end{subfigure}
\begin{subfigure}{.25\textwidth}
	\centering	\includegraphics[width=1\linewidth]{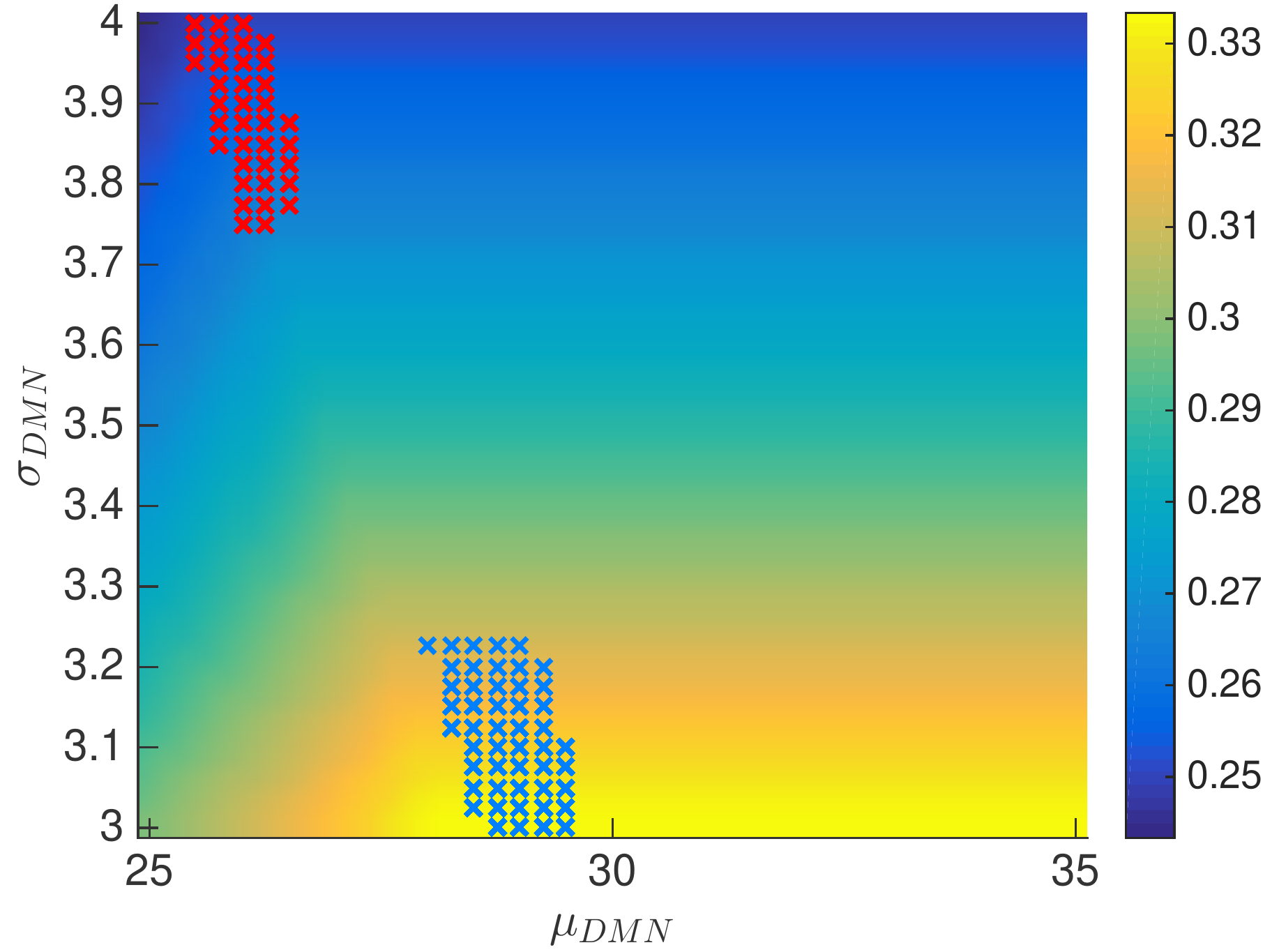} 
\end{subfigure}%
\begin{subfigure}{.25\textwidth}
	\centering \includegraphics[width=1\linewidth]{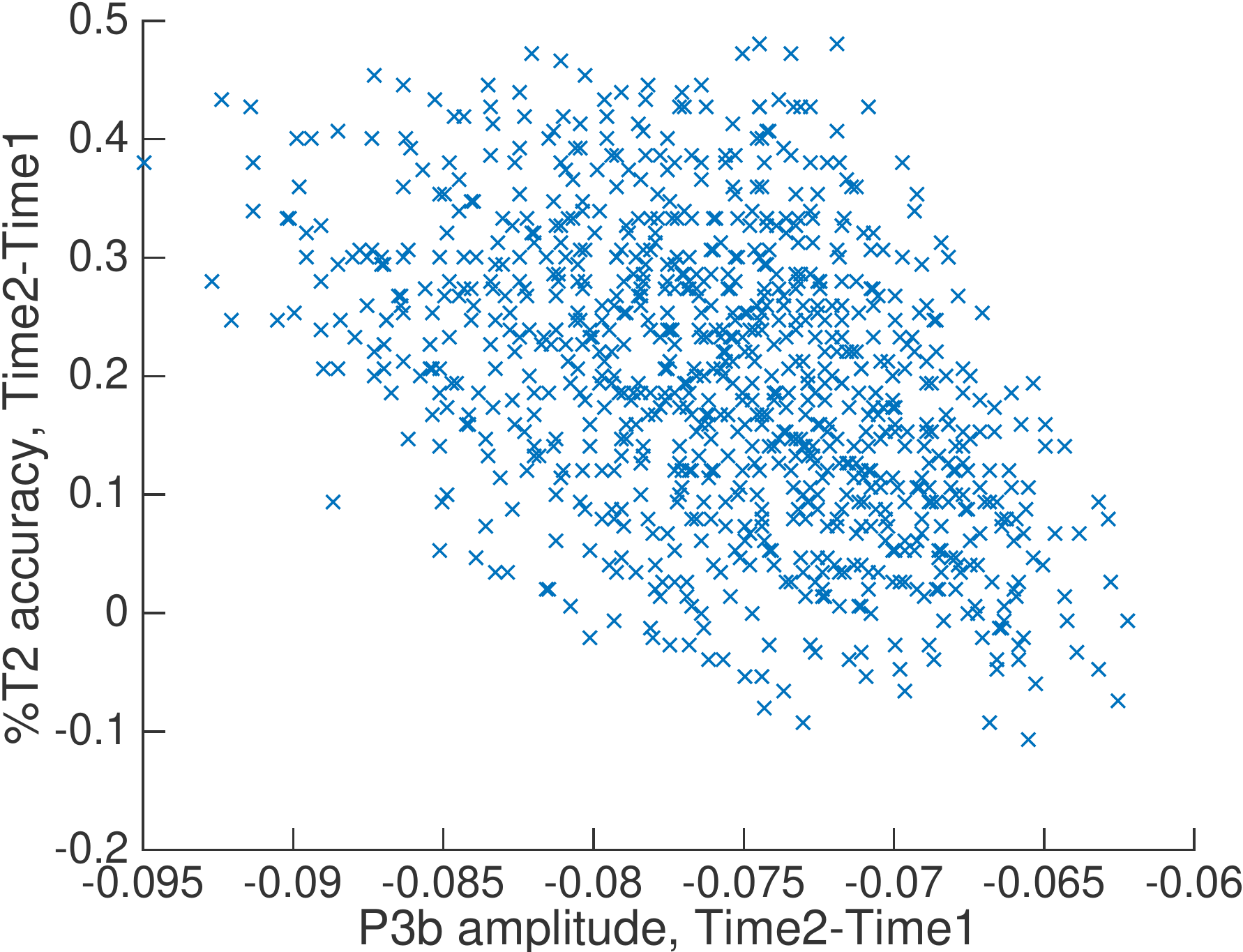}
\end{subfigure}
\caption{{\bf Model behavior in DMN parameter space}
Top left: lag-1 sparing probability as a function of DMN activity parameters (mean and variance). The color indicates the probability of crossing the blinking threshold at lag 2 but not at lags 1 and 5.
Top right: T2 detection probability for a T2-T1 lag of 4. Bottom left: The model P3b amplitude defined in Eq.~\ref{eq:P3b_model}, for a T2-T1 lag of 4. Blue and red crosses correspond to possible $(\mu_{DMN},\sigma_{DMN})$ values at time 1 and time 2 respectively, for which the model reproduces lag-1 sparing as well as the main findings of \cite{Slagter2007}, namely an increase of T2 detection probability from about 60\% to about 80\% (ibid. figure 2) and a reduction of about 25\% in P3b amplitude (ibid. figure 3) and.
Bottom right: Change in T2 detection accuracy as a function of the change in P3b for $(\mu_{DMN},\sigma_{DMN})$ pairs reproducing lag-1 sparing and the effects of meditation (cf. figure 4. in \cite{Slagter2007}).}
\label{fig:phaseSpace}
\end{figure}

 
Next, we tested whether the effects of mental training on AB performance and the corresponding P3b ERP can be reproduced by the model. In \cite{Slagter2007} it was shown that mental training, in the form of a 3 month mindfulness meditation retreat, resulted in significantly lower blinking probability and larger reduction in T1 elicited P3b amplitudes for non-blinking vs blinking trials in expert meditators compared to a group of matched novice controls who did not attend the retreat. Furthermore, the meditators' reduction in T1 evoked P3b amplitudes was significantly correlated with improvement in T2 detection accuracy, suggesting that the ability to detect T2 depends on efficient deployment of attentional resources to T1 \cite{Slagter2007}. Using the theoretical P3b amplitude defined in Eq. \ref{eq:P3b_model}, the model was able to reproduce the three-way interaction (meditators vs. novices, before vs. after retreat, blinking vs. non-blinking trials), Fig. \ref{fig:meditationEffect}, (cf. figure 3 in \cite{Slagter2007}), and the relationship between T2 detection accuracy and reduction in T1 elicited P3b amplitude. Fig. \ref{fig:phaseSpace} bottom right, (cf. figure 4 in \cite{Slagter2007}). 
\par
The model reproduced these results only for certain values of DMN activity levels before and after training. We thus wanted to identify all pairs of $(\mu_{DMN},\sigma_{DMN})$ points, representing DMN activity at times 1 and 2, which reproduce the main findings reported in \cite{Slagter2007}, namely a reduction of about 25\% in T1 elicited P3b amplitude and an increase of about 20\% (from 60\% to 80\%) in T2 detection accuracy. In addition to reproducing these empirical findings, we required that these points reproduce the lag-1 sparing effect, which we defined as the condition of crossing the blinking threshold for T2-T1 lags of 2 but not for lags 1 or 5 and higher. We plotted the probability of lag-1 sparing occurrence, as well as the model P3b amplitude (Eq.~\ref{eq:P3b_model}) and T2 detection probability, both at a T2-T1 lag of 4 (Fig.~\ref{fig:phaseSpace}, top left, bottom left and top right respectively). We then tested which pairs of $(\mu_{DMN},\sigma_{DMN})$ points, representing DMN activity at time 1 and 2, yield a substantial probability ($>0.2$) of lag-1 sparing occurrence at both times and reproduce the main empirical findings of \cite{Slagter2007} mentioned above (Fig. \ref{fig:phaseSpace}, blue and red crosses, top left and right and bottom left). For these pairs of points, we plotted the change in T2 detection probability a function of the reduction in P3b amplitude (Fig. \ref{fig:phaseSpace}, bottom right). This yielded a set of DMN activity parameter pairs which reproduce the empirical correlation between improvement in T2 detection accuracy and reduction in T1 elicited P3b amplitude as reported in figure 4 of \cite{Slagter2007}. Thus, the model provides a quantifiable relationship between DMN activity levels, AB task performance and T1 evoked P3b amplitudes, which can presumably be tested empirically.
\section{Discussion}
    \subsection{Meditation, P3b and the default mode network}
  Differences in attentional processing between expert meditators and novices have been reported in several studies over the last years (for an overview see \cite{Lutz2009a}). Of particular interest here is a study dealing with the effect of meditation on the AB and the related P3b potentials evoked during the task performance \cite{Slagter2007}. The P3b wave is an event related potential (ERP) component linked with attention allocation and memory encoding of targeted stimuli \cite{Wickens1983,Polich2007}. The results reported in \cite{Slagter2007} show that experienced meditators exhibit a smaller attentional blink effect (higher T2 recognition accuracy), and larger amplitudes of T1 evoked P3b compared to a matched control group of meditation novices. In addition, after attending a 3 month intensive meditation retreat, the meditators showed a significant reduction of T1 evoked P3b amplitudes but only during non-blinking trials. The novice control group, who did not attend the meditation retreat, showed a significantly smaller reduction in both attentional blink size and T1 evoked P3b amplitude. The magnitude of reduction in T1 evoked P3b amplitudes during no-blink trials was correlated with the improvement in T2 recognition accuracy for both meditators and novices. These results suggest that the ability to identify T2 depends on efficient attentional resource allocation to T1 and that mental training, such as mindfulness meditation, can improve this resource allocation process. The model suggests a concrete mechanism, namely the reduction in DMN noise levels, through which this process can take place. This also provides a quantitative interpretation of the widespread claim that mindfulness meditation reduces ongoing mental noise in the brain. 
  \par
  Interestingly, the T1 evoked P3b amplitudes were larger for the experienced meditators compared to the novices in all experimental conditions (blinking and non-blinking trials, before and after the retreat), with the possible exception of non-blinking trials after the retreat, where they were of comparable magnitude (figure 3. in \cite{Slagter2007}). The model suggests that these differences may be due to smaller fluctuations (noise variance) in the DMN activity of experienced meditators compared to novices.
  \par
  Over the last few years, several studies have shown that different types of meditation are linked with reduced DMN activity levels compared to rest \cite{Brewer2011,Pagnoni2012} and other active tasks \cite{Garrison2015}. It has also been shown recently that experienced mindfulness meditators exhibit reduced resting state DMN activity and fluctuations amplitudes during a visual memory task \cite{Berkovich-Ohana2016} compared to novice controls. By proposing that DMN activity reduction is the neural mechanism by which meditation increases attentional capacity, our model relates these studies with the finding that meditation affects performance and brain potentials in the AB task \cite{Slagter2007,Lutz2009a,VanVugt2014}.
  \par 
  In a 2004 study measuring BOLD activation levels during AB task performance \cite{Marois2004}, the only brain area reported to exhibit decreased activation for non-blinking vs. blinking trials was the right temporoparietal junction (TPJ), a DMN associated region known to deactivate during attention demanding tasks \cite{Shulman2007}. Other, non DMN related regions, showed the opposite effect, namely increased activation during non-blinking compared to blinking trials \cite{Marois2004}. This finding is also in line with the model's hypothesis regarding the relationship between DMN activity and blinking probability.
\subsection{Comparison with existing models}
Several mathematical models for the AB effect have been suggested over the past decade or so (for an overview see the section ``Formal Theories" in \cite{Dux2009rev}). Formally, these models can be divided into two classes: connectionist and symbolic \cite{Minsky1991}. Connectionist, also known as neural-network models (e.g \cite{Wyble2009,Nieuwenhuis2005,Chartier2004,Fragopanagos2005,Olivers2008}), usually rely on tuning a large number of parameters without providing much insight regarding the underlying mechanisms. On the other hand, symbolic, or computationalist models (e.g. \cite{Shih2007,Taatgen2009}), are often described using complex box-and-arrow type rules which seem rigid and ad-hoc. Another way of categorizing AB models is by the underlying cause for blinking. The two pravelant mechanisms are capacity limitations and attentional control (\cite{Martens2010}). Models emphasizing capacitiy limits attribute blinking to some resource bottleneck at the attentional or working memory systems, while those emphasizing a control mechanism posit that blinking is due to top-down inhibition of attention while T1 is being processed.
\par
The model proposed here can be broadly categorized as symbolic and capacity limited. However, it is based on first-principles (linear dynamical-systems) and is essentially described by a single equation (Eq.~\ref{eq:model}) assuming only a few free parameters. The model also incorporates a top-down control element by modulating the available attentional capacity through changes in DMN activity. As discussed below, this element enables the model to explain certain findings which challenge the notion of a purely capacity-limited account of the AB, such as the``spreading'' of lag-1 sparing (see below).
\par
While simple, the model can explain many findings from the AB literature and, in particular, reproduce the results reported in \cite{Slagter2007} relating mental training with improved AB performance and reduced P3b amplitudes. We discuss below the central findings relating to the AB and describe how they are accounted for by the model. 
\subsubsection{Reversal of T1 and T2}
It has been reported that when T2 appears immediately after T1 (lag 1 trials), identification of T2 is often superior to T1 and report order is often reversed \cite{Hommel2005,Chun1995}. The model is consistent with these effects since it posits that during the cognitive representation stage, T1 may be partially occluded by T2 and both are merged together into a single ``cognitive'' trace. This process of occlusion and merging may explain degredation of T1 detection probability and T1,T2 order confusion.          
\subsubsection{Spreading of the sparing}
This refers to a set of findings showing that AB can be attenuated or even eliminated as long as a target is not followed by a distractor. Thus, when subjects were presented with RSVP streams containing three consecutive targets (T1,T2,T3), there was no deficiency in reporting T3 even though it appeared at the temporal position in which blinking is typically maximal (lag 2) \cite{DiLollo2005}. This phenomenon was called ``spreading of lag-1 sparing'' \cite{Olivers2007} since the attenuation of blinking at lag-1 spreads to additional (target) stimuli. A related finding is the attenuation of AB when subjects were required to give a whole report of a six letter sequence, i.e. when all stimuli were considered targets, compared to the standard case in which only two letters had to be reported \cite{Nieuwenstein2006}. Such findings pose a challenge for limited-capacity accounts of the AB and seem to indicate the workings of a top-down attentional control mechanism \cite{Martens2010}. However, the model presented here can provide an interesting explanation of these findings by assuming that a consecutive sequence of targets temporarily reduces DMN fluctuations, perhaps by focusing the subject's attention on the task, thus reducing blinking probability until appearance of the next distractor. This hypothesis explains ``spreading'' effects in terms of an interaction between capacity limitation (the blinking threshold), and a top-down control mechanism (temporary reducing DMN noise). The partial occlusion of earlier by later targets at the cognitive representation stage may also explain the poorer identification of T1 typically observed when three targets are presented sequentially \cite{DiLollo2005,Kawahara2006}. 
\subsubsection{No T2 evoked P3b during blinking trials}
In \cite{Vogel1998} it was shown that missed T2 targets do not evoke a P3b ERP. In the model, these trials correspond to cases in which the attentional resources were depleted by the T1, or the combined representation of T1 and T2, signal. This results in a large P3b amplitude elicited by the T1, or combined T1,T2, signal and no attentional resource capture, and thus no P3b signal, evoked in response to T2 itself.
\subsubsection{Distractor and target saliency}
The experiments reported in \cite{Folk2002,Maki2006} showed that salient distractors which match features with the target set (having the same color), can also trigger an AB for a subsequent target. The model can account for these findings by representing salient distractors as pulses with smaller amplitudes (compared to targets) at the sensory signal level. These distractor pulses will also contribute towards triggering an AB but to a lesser extent than an additional target pulse would. Using a similar amplitude modulation mechanism, the model can explain why targets which capture the attention more powerfully (e.g. by switching their color compared to pre-target distractors) have a higher propensity for causing AB \cite{Dux2008}.
\subsubsection{Effect of inter-target blanks}  
There are mixed results in the AB literature regarding the effects of inserting a blank stimulus between T1 and T2. In some cases blinking was reported although at attenuated levels compared to the case of a  (non-blank) distractor (e.g.~\cite{Raymond1992,Chua2005,Visser2007}), while other studies showed no blinking when inserting inter-target blanks (e.g.~\cite{Chun1995,Grandison1997,Breitmeyer1999}). Such blanks can be represented in the model by small inter-target impulses whereas (non-blank) distractors can be represented by somewhat larger, yet still smaller than target, pulses. While this particular analysis is beyond the scope of the current work, we wish to point out that this and similar questions can be addressed by the model using slight modifications of the basic framework. 
\subsubsection{Task-irrelevant mental activity}
A somewhat counterintuitive series of findings reported in \cite{Olivers2005,Arend2006} suggest that certain seemingly task irrelevant stimuli, or mental activities (background music or visual motion, performing a concurrent memory task etc.) can significantly attenuate the AB. These interventions presumably load the attentional channel, and thus seem at odds with a capacity limited account of the AB. However, the model provides an interesting explanation for these findings by hypothesizing that such activities may reduce DMN activity levels, and this increase attentional capacity, perhaps by focusing subjects on their immediate environment thus reducing mind wandering. 
\section{Conclusion}
This paper proposed a model of attentional blink which is simple and parsimonious while providing explanatory and predictive power. The model's main features are a dynamical system's account of attentional capacity with a top-down control mechanism in the form of DMN activity, represented by stochastic noise loading the system's capacity. The model generates quantifiable predictions relating DMN activity levels, AB task performance and the P3b ERP, which can be tested in future experiments. 
\section{Acknowledgments}
The authors thank the ICRI-CI, the Israel Science Foundation, the Gatsby Charitable Foundation and the European Research Council.
We wish to thank Leon Deuoell and Tamar Regev for helpful discussions.
\bibliographystyle{IEEEtran}
\bibliography{Attentional_Blink}

\end{document}